\documentclass{elsart}

\usepackage{graphicx}
\usepackage{latexsym}
\usepackage{amssymb,amsmath,latexsym}

\begin{document}

\begin{frontmatter}


\title{On the 4D effective theory in warped compactifications 
with fluxes and branes}
%
\author{Kazuya Koyama}$^{1}$,
\ead{Kazuya.Koyama@port.ac.uk}
\author{Kayoko Koyama}$^{2}$,
\ead{kk44@sussex.ac.uk}
\author{Frederico Arroja}$^{1}$
\ead{Frederico.Arroja@port.ac.uk}
\address{1. Institute of Cosmology and Gravitation, University of
Portsmouth, Portsmouth PO1 2EG, UK}
\address{2. Department of Physics and Astronomy, University of Sussex, 
Brighton,\\ BN1 9QH, UK}
%
%
%
%
%
\begin{abstract}
We present a systematic way to derive the four-dimensional effective 
theories for warped compactifications with fluxes and branes 
in the ten-dimensional type IIB supergravity. The ten-dimensional 
equations of motion are solved using the gradient expansion method 
and the effective four-dimensional equations of motions are derived 
by imposing the consistency condition that the total derivative terms
with respect to the six-dimensional internal coordinates vanish when 
integrated over the internal manifold. By solving the effective 
four-dimensional equations, we can find the gravitational backreaction
to the warped geometry due to the dynamics of moduli fields, branes 
and fluxes. 
\end{abstract}
%
%
%
%
%
%
%
%
%
%
\end{frontmatter}
%
%

\section{\label{sec:INTRO}Introduction}
Warped string compactifications with fluxes and branes have provided a novel
approach to long standing problems of particle physics 
and cosmology. Inspired by the earlier work by 
Randall and Sundrum \cite{MAY},
Giddings \emph{et al} (GKP) \cite{Giddings:2001yu} showed that, in
type IIB string theory, it is possible to realize the warped
compactifications that can accommodate the large hierarchy between 
the electroweak scale and the Planck scale. In the GKP model, the
warping of the extra-dimensions is generated by the fluxes and the
presence of D-branes and orientifold planes. All moduli fields are 
stabilized due to fluxes except for the universal K\"ahler modulus.
The warped compactifications
also provide a promising background for cosmological inflation. 
An inflaton can be identified with the internal coordinate of a mobile
D3 brane moving in a warped throat region \cite{KKLMMT}. 

So far, most works are essentially based on effective 4D theories 
derived by a dimensional reduction. However, it is a non-trivial 
problem to derive the 4D effective theories including the warping 
and branes.
In fact there have been debates on the validity of the derivation 
of the 4D effective theories. In a conventional Kaluza-Klein theory, 
the effective theory can be derived by assuming a factorizable 
ansatz for the higher-dimensional fields. This approach has been 
applied even in the presence of the warping. For example, the 
10D metric ansatz of the form
\begin{equation}
ds_{10}^2 = h^{-1/2}(y) e^{-6 u(x)} g_{\mu \nu}(x) dx^{\mu} dx^{\nu}
+ h^{1/2}(y) e^{2 u(x)} \gamma_{mn} dy^m dy^n,
\label{wrong}
\end{equation}
is often used where $x$ denotes the coordinates of the 4D non-compact spacetime 
and $y$ denotes the coordinates of the 6D compact manifold. The 
function $u(x)$ is identified with the universal K\"ahler 
modulus. 
However, it has been criticized that the higher-dimensional 
dynamics do not satisfy this ansatz \cite{deAlwis}. Thus it is 
required to derive the 4D effective theory by starting from a correct 
ansatz and consistently solving the 10D equations of motion. 

These attempts were initiated in Refs.~\cite{HK1,Giddings:2005ff,HK2}. 
Ref~\cite{HK1} derived an exact time-dependent 
10D solution that describes the instability of the warped compactification 
due to the non-stabilized K\"ahler modulus. It was shown 
that this dynamical solution cannot be described by the 
metric ansatz (\ref{wrong}). Ref.~\cite{Giddings:2005ff} derived 
the potential for the moduli fields by consistently solving the 10D equations 
of motion using the metric ansatz that is consistent with the dynamical 
solutions in the 10D theory. It was found that the universal K\"ahler
modulus is not a simple scaling of the internal metric as is assumed in 
Eq.~(\ref{wrong}). 
A similar consistent ansatz was proposed in Ref~\cite{HK2} and the 4D effective 
theory without potentials was derived. 

In this letter, we build on these works 
to present a systematic way to 
derive the 4D effective theory by consistently
solving the 10D equations of motion. We exploit the so-called gradient 
expansion method, which has been shown to be a powerful method to derive 
the 4D effective theory in the context of 5D brane world models 
\cite{gradient}.
Our method also provides a scheme to study the gravitational backreaction
to the warped geometry due to the moduli dynamics, branes and fluxes.

\section{\label{sec:AAE}10D equations of motion}
Let us start by describing the type IIB supergravity based on 
Ref.~\cite{Giddings:2001yu}.
In the Einstein frame, the bosonic part of the action for the type IIB supergravity is
given by
\begin{eqnarray}
S_{IIB} &=& \frac{1}{2\kappa_{10}^2}\int
d^{10}x\sqrt{-g}\left(R-\frac{\partial_M\tau\partial^M\bar{\tau}}{2\left(\mbox{Im}
\tau\right)^2}-\frac{G_{(3)}\cdot\bar{G}_{(3)}}{12\mbox{Im}\tau}-\frac{\tilde{F}_{(5)}^2}{4\cdot5!}\right) \nonumber\\
&&-\frac{i}{8\kappa_{10}^2}
\int\frac{C_{(4)}\wedge G_{(3)}\wedge\bar{G}_{(3)}}{\mbox{Im}\tau},
\label{10DACTION}
\end{eqnarray}
where the combined 3-form flux is $G_{(3)}=F_{(3)}-\tau H_{(3)}$
with $F_{(3)}=dC_{(2)}$, $H_{(3)}=dB_{(2)}$ and
\begin{equation}
\tau=C_{(0)}+ie^{-\phi},
\end{equation}
where
$\phi$ is the so-called dilaton, $C_{(j)}$ is the Ramond-Ramond potential of rank $(j)$ and $B_{(2)}$ is the NS-NS potential. The 5-form field is given by 
\begin{equation}
\tilde{F}_{(5)}=F_{(5)}-\frac{1}{2}C_{(2)}\wedge
H_{(3)}+\frac{1}{2}B_{(2)}\wedge F_{(3)},
\end{equation}
with $F_{(5)}=dC_{(4)}$.
The total action in our model is
\begin{equation}
S=S_{IIB}+S_{loc},\label{10DACTIONTOTAL}
\end{equation}
where the term $S_{loc}$ is the action for the localized sources such as 
D3-branes and O3 planes;
\begin{equation}
S_{loc}=\sum_j \left(-\mu_3\int d^4x\sqrt{-g_j}+T_3\int d^4xC_4\right),
\end{equation}
where the integrals are calculated over the 4D non-compact space at the point $j$ in the compact space and $g_j$ is the determinant of the induced metric on a 
brane at the point $j$ ($\mu_3$ is positive/negative for D3-branes/O3 planes).

We can derive the equations of motion for the fields from the 
action (\ref{10DACTION}). The trace reversed Einstein equations are given by
\begin{eqnarray}
R_{AB}&=&\frac{\mbox{Re}\left(\partial_A\bar{\tau}\partial_B\tau\right)}{2\left(\mbox{Im}\tau\right)^2}+\frac{1}{4\mbox{Im}\tau}\left[\mbox{Re}\left(G_{ACD}\bar{G}_B^{\;\;\;CD}\right)-\frac{1}{12}G_{CDE}\bar{G}^{CDE}{}^{(10)}g_{AB}\right]\nonumber\\
&&+\frac{1}{96}\tilde{F}_{AP_1...P_4}\tilde{F}_B^{\;\;P_1...P_4}
+\kappa_{10}^2\left(T_{AB}^{loc}-\frac{1}{8}{}^{(10)}g_{AB}T^{loc}\right),\label{10DEE}
\end{eqnarray}
where $T_{AB}^{loc}$ is the energy-momentum tensor for the localized sources defined by
\begin{equation}
T_{AB}^{loc}=-\frac{2}{\sqrt{-{}^{(10)}g}}\frac{\delta S_{loc}}{\delta {}^{(10)}g^{AB}},
\end{equation}
and ${}^{(10)}g_{AB}$ is the metric for the 10D spacetime.

Following Refs.~\cite{Giddings:2005ff,HK2}, we take the 10D metric as 
\begin{equation}
dS_{10}^2= h^{-\frac{1}{2}}(x,y)g_{\mu\nu}(x,y)dx^\mu
dx^\nu + h^{\frac{1}{2}}(x,y)\gamma_{mn}(y)dy^mdy^n,
\label{metric}
\end{equation}
with
\begin{equation}
h(x,y)=h_1(y)+h_0(x),
\end{equation}
where upper case latin indices run from 0 to 9, greek indices run
from 0 to 3 (non-compact dimensions) and lower case latin indices
run from 4 to 9 (compact dimensions). In the following, all indices 
are raised by $g_{\mu \nu}$ and $\gamma_{m n}$. The function $h_0(x)$ is the 
so-called universal K\"ahler modulus. It should be emphasized that this is not 
a simple scaling of the internal metric. 

Using our metric ansatz, we can calculate the 10D Ricci tensor.
The mixed component is calculated as 
\begin{equation}
R_{\mu p}=- g^{\alpha\beta} K_{\mu\beta
p|\alpha}+K_{p,\mu}-\frac{1}{2}h^{-1}h_{,\mu} K_p,
\end{equation}
where we defined 
\begin{equation}
K_{\mu\nu p}\equiv-\frac{1}{2} g_{\mu\nu,p}, \quad K_p\equiv g^{\mu\nu} K_{\mu\nu p},
\end{equation}
and $|$ denotes the covariant derivative with respect to
$g_{\mu\nu}$, that is,
\begin{equation}
K_{\mu\delta p|\alpha}\equiv K_{\mu\delta
p,\alpha}-{}^{(4)}\Gamma_{\mu\alpha}^\sigma K_{\sigma\delta
p}- {}^{(4)}\Gamma_{\delta\alpha}^\sigma K_{\mu\sigma p},
\end{equation}
where ${}^{(4)} \Gamma_{\mu\alpha}^\sigma$ is the Christoffel symbol
constructed from $g_{\mu \nu}$.
The non-compact components are given by
\begin{eqnarray}
R_{\mu\nu}&=&{}^{(4)}R_{\mu\nu}(g)-h^{-1}h_{|\mu\nu}+\frac{1}{4}h^{-1} g_{\mu\nu}h_{|\alpha}^{|\alpha}+
\frac{1}{4}h^{-2} g_{\mu\nu}h_{;a}^{;a}-\frac{1}{4}h^{-3}h_{;a}h^{;a} g_{\mu\nu}
\nonumber\\ &&+h^{-1} K_{\mu\nu
b}^{\;\;\;;b}-\frac{1}{4}h^{-2}h^{;b} K_b g_{\mu\nu}-h^{-1} K_{\mu\nu}^{\;\;b}
K_b
+
2h^{-1} K_{\nu\;b}^{\;\delta} K_{\mu\delta}^{\;\;\;b},
\end{eqnarray}
where $;$ denotes covariant derivative with respect to
$\gamma_{ab}$, that is,
\begin{equation}
K_{\mu\nu b;a}\equiv K_{\mu\nu
b,a}-{}^{(6)}\Gamma_{ab}^c K_{\mu\nu c},
\end{equation}
where ${}^{(6)}\Gamma_{ab}^c$ is the Christoffel symbol constructed 
from $\gamma_{ab}$.
The compact components of the Ricci tensor are
\begin{eqnarray}
R_{ab}&=&{}^{(6)} R_{ab}(\gamma)+\frac{1}{4}h^{-2}h_{;d}h^{;d} \gamma_{ab}-\frac{1}{4}h^{-1}h_{;d}^{;d} \gamma_{ab}
-\frac{1}{4}h_{|\delta}^{|\delta} \gamma_{ab}-\frac{1}{2}h^{-2}h_{;a}h_{;b}
\nonumber\\
&&+K_{a;b}-\frac{1}{2}h^{-1}\left(h_{,b} K_a+h_{,a} K_b\right)+\frac{1}{4}h^{-1}h^{;c} K_c \gamma_{ab}
-K^{\alpha\beta}_{\;\;\;a} K_{\alpha\beta b}.
\end{eqnarray}
Note that 
the self duality of the 5-form field must be imposed by hand
\begin{equation}
\tilde{F}_{(5)}=\ast \tilde{F}_{(5)}.\label{F5SD}
\end{equation}
We shall take the self-dual 5-form field in the form
\begin{equation}
\tilde{F}_{(5)}=\left(1+\ast\right)\sqrt{-g}d_y\alpha(x,y)
\wedge dx^0\wedge dx^1\wedge dx^2\wedge dx^3,\label{F5}
\end{equation}
Then the self duality condition (\ref{F5SD}) is automatically satisfied 
and the only non-zero components of $\tilde{F}_{(5)}$ are
\begin{equation}
\tilde{F}_{(5)a\mu_0\cdots\mu_3}=\alpha_{;a} {}^{(4)} 
\varepsilon_{\mu_0\cdots\mu_3},
\end{equation}
\begin{equation}
\tilde{F}_{(5)n_1\cdots
n_5}=-h^2 \alpha^{;c} {}^{(6)} \varepsilon_{cn_1\cdots n_5},
\end{equation}
where ${}^{(4)}\varepsilon_{\mu_0\cdots\mu_3}$ denotes the Levi-Civita
tensor with respect to $g_{\mu\nu}$ and
${}^{(6)}\varepsilon_{cn_1\cdots n_5}$ denotes the Levi-Civita tensor with 
respect to $\gamma_{mn}$. 

\section{\label{sec:GEM}Gradient expansion method}
In this section, we will use the gradient expansion method 
to solve the 10D Einstein equations
(\ref{10DEE}).
\subsection{\label{subsec:Approx}Gradient expansion}
The gradient expansion is based on the assumption that 
$x$ derivatives are suppressed compared with $y$ derivatives
\begin{equation}
\partial_x^2 \ll \partial_y^2.
\end{equation}
Using this assumption, we can reduce the partial differential 
equations with respect to $x$ and $y$ to a set of ordinary differential 
equations with respect to $x$. 
We expand the metric as 
\begin{equation}
g_{\mu\nu}(x,y)=\stackrel{(0)}{g_{\mu\nu}}(x,y)+\stackrel{(1)}{g_{\mu\nu}}(x,y)+\cdots,
\end{equation}
where the first order quantities are of the order $\partial_x^2/\partial_y^2$.
Accordingly, $K_{\mu \nu p}$ is also expanded as 
\begin{equation}
K_{\mu\nu p}=\stackrel{(0)}{K_{\mu\nu
p}}+\stackrel{(1)}{K_{\mu\nu p}}+\cdots.
\end{equation}

\subsection{\label{subsec:zero}First order equations}
We assume that at zeroth-order, the axion/dilaton is constant
and the 3-form field only has non-zero components in the compact space, i.e.
\begin{equation}
G_{(3)}=\frac{1}{3!}G_{abc}(y)dy^a\wedge dy^b\wedge dy^c.
\end{equation}
Furthermore, we assume that the zeroth-order metric is independent of the 
compact coordinates
\begin{equation}
\stackrel{(0)}{g_{\mu\nu}}(x,y)=\stackrel{(0)}{g_{\mu\nu}}(x).\label{0metric}
\end{equation}

The Bianchi identity/equation of motion for the 5-form field 
becomes 
\begin{equation}
\alpha_{;c}^{;c} 
+ 2 h^{-1} h_{;c} \alpha^{;c}
=ih^{-2}\frac{G_{pqr}\ast_6\bar{G}^{pqr}}{12\mbox{Im}\tau}+2\kappa_{10}^2h^{-2} T_3\rho_3^{loc},\label{F5Bi2}
\end{equation}
at zeroth order where $\ast_6$ is the Hodge dual with respect to $g_{ab}$ and we defined 
a rescaled D3 charge density $\rho_3^{loc}$ which does not depend on $h$.
We define $\omega$ as 
\begin{equation}
\omega = \alpha -h^{-1}.
\end{equation}
Then the Bianchi identity is rewritten as 
\begin{equation}
-h_{;c}^{;c} = - h^2 \omega^{;c}_{;c} - 2 h h^{;c} \omega_{;c} 
+i \frac{G_{pqr}\ast_6\bar{G}^{pqr}}{12\mbox{Im}\tau}
+2\kappa_{10}^2  T_3\rho_3^{loc}.
\end{equation}
The non-compact components of the Einstein equations (\ref{10DEE}) can 
be rewritten as 
\begin{eqnarray}
\!\!\!\!\!\!{}^{(4)}R_{\mu\nu} && - h^{-1}\left(h_{|\mu\nu}-\frac{1}{4}\stackrel{(0)}{g_{\mu\nu}}h_{|\alpha}^{|\alpha}\right) +h^{-1}\stackrel{(1)}{K_{\mu\nu
b}^{\;\;\;;b}}-\frac{1}{4}h^{-2}h^{;b}\stackrel{(1)}{K_b}\stackrel{(0)}{g_{\mu\nu}}
+ \frac{1}{4} 
\omega^{;c}_{;c} \stackrel{(0)}{g_{\mu \nu}}
\nonumber\\ &&
=-\frac{h^{-2}}{96\mbox{Im}\tau}\stackrel{(0)}{g_{\mu\nu}}\left|iG_{(3)}-\ast_6G_{(3)}\right|^2+
\frac{\kappa_{10}^2}{2}h^{-2}\stackrel{(0)}{g_{\mu\nu}}\left(T_3\rho_3^{loc}-\mu_3(y)\right)
,\label{1stmunu}
\end{eqnarray}
where ${}^{(4)} R_{\mu\nu}$ denotes the Ricci tensor constructed from $g_{\mu \nu}$,
$\mu_3(y)= \mu_3 \delta(y-y_i)/\sqrt{\gamma}$ which is independent 
of $h$ and we dropped the non-linear term in $\omega$.
In the same way, the compact equations can be rewritten as 
\begin{eqnarray}
\!\!\!\!\!\!{}^{(6)}R_{ab} &-& \frac{1}{4}h_{|\delta}^{|\delta} \gamma_{ab}
+\stackrel{(1)}{K_{a;b}}-\frac{1}{2}h^{-1}\left(h_{,b}\stackrel{(1)}{K_a}+h_{,a}\stackrel{(1)}{K_b}\right)+\frac{1}{4}h^{-1}h^{;c}\stackrel{(1)}{K_c}\gamma_{ab} \nonumber\\
&=&  \frac{1}{4} h \omega^{;c}_{;c} \gamma_{ab}+ \frac{1}{2}  
\left(h_{;a} \omega_{;b} +h_{;b} \omega_{;a}   \right) \nonumber\\
&& +\frac{h^{-1}}{4\mbox{Im}\tau}\left[\mbox{Re}\left(G_{acd}\bar{G}_b^{cd}\right)-\frac{1}{12}G_{cde}\bar{G}^{cde}\gamma_{ab}-\frac{i}{12}
G_{pqr}\ast_6\bar{G}^{pqr}\gamma_{ab}\right] \nonumber\\
&& +\kappa_{10}^2\left(T_{ab}^{loc}-\frac{1}{8}
h^{\frac{1}{2}} \gamma_{ab}T^{loc}-\frac{1}{2}h^{-1} T_3
\rho_3^{loc} \gamma_{ab}\right), \label{1stab}
\end{eqnarray}
where ${}^{(6)} R_{ab}$ denotes the Ricci tensor constructed from 
$\gamma_{a b}$.

The GKP solution is obtained by taking
\begin{equation}
\ast_6G_{(3)}=iG_{(3)} \label{ISD}, \quad \omega=0,
\end{equation}
with local sources that satisfy $\mu_3(y)=T_3 \rho_3^{loc}$. 
With these conditions, it is straightforward to show that 
the 10D equations motion are satisfied. 
Then the warp factor is determined by 
\begin{equation}
-h^{;a}_{;a} = \frac{1}{12 \mbox{Im} \tau} G_{pqr} 
\bar{G}^{pqr} + 2 \kappa_{10}^2 T_3 \rho_3^{loc}.
\label{h1}
\end{equation}
Our strategy is to assume that $\omega$, ${}^{(6)} R_{ab}$ and the right hand 
sides of Eqs.~(\ref{1stmunu}) and (\ref{1stab}) are first 
order in the gradient expansion and take into account these contributions 
as a source for the 4D dynamics of $g_{\mu \nu}(x)$ and $h_0(x)$.

\section{4D Effective equations}
In this section we solve the 10D Einstein equations to get the 
4D effective Einstein equations. Hereafter we omit the superscript $(i)$
that denotes the order of the gradient expansion. 
A key point is the consistency condition that implies that the integration of the 
total derivative term over the 6D internal dimension vanishes
\begin{equation}
\int  d^6 y \sqrt{\gamma} v_{;c} =0,
\end{equation}
for an arbitrary $v$.
This condition provides the boundary conditions for the 10D gravitational fields 
and yields the 4D effective equations. 

First let us take the trace of Eq.~(\ref{1stmunu});
\begin{equation}
{}^{(4)}R+\left(h^{-1} K_c  + \omega_{;c}  \right)^{;c}
=-\frac{h^{-2}}{24\mbox{Im}\tau}\left|iG_{(3)}-\ast_6G_{(3)}\right|^2 
+2\kappa_{10}^2h^{-2}\left(T_3\rho_3^{loc}-\mu_3(y)\right).
\label{tracemunu}
\end{equation}
Then integrating Eq.~(\ref{tracemunu}) over the internal manifold, we get 
\begin{eqnarray}
{}^{(4)} R &=& -\frac{1}{24V_{(6)}\mbox{Im}\tau}\int d^6y\sqrt{\gamma}h^{-2}\left|iG_{(3)}-\ast_6G_{(3)}\right|^2 \nonumber\\
&& +2\frac{\kappa_{10}^2}{V_{(6)}}\int d^6y\sqrt{\gamma}h^{-2}\left(T_3\rho_3^{loc}-\mu_3(y)\right),
\end{eqnarray}
where $V_{(6)}$ is the volume of the internal space
\begin{equation}
V_{(6)}\equiv \int d^6y\sqrt{\gamma}.
\end{equation}

On the other hand, by combining Eq.~(\ref{1stmunu}) with 
Eq.~(\ref{tracemunu}), we obtain the traceless part of the equation
\begin{equation}
{}^{(4)}R_{\mu\nu}-\frac{1}{4} g_{\mu\nu} {}^{(4)}R-
h^{-1}\left(h_{|\mu\nu}-\frac{1}{4} g_{\mu\nu} h_{|\delta}^{|\delta} \right)+
h^{-1} \left( K_{\mu\nu b}   
-\frac{1}{4} g_{\mu \nu} K_b \right)^{;b}=0.
\label{Sigma}
\end{equation}
Integrating this over the compact space we obtain
\begin{equation}
{}^{(4)} R_{\mu\nu}-\frac{1}{4}g_{\mu\nu} {}^{(4)}R=H^{-1}\left(h_{|\mu\nu}-\frac{1}{4}
g _{\mu\nu} h_{|\delta}^{|\delta}\right),
\label{Meom}
\end{equation}
where the function $H(x)$ is defined as
\begin{equation}
H(x)\equiv h_0(x)+C,
\end{equation}
where
\begin{equation}
C \equiv \frac{1}{V_{(6)}}\int d^6y\sqrt{\gamma}h_1(y).
\end{equation}
Here $h_1(y)$ is obtained by solving Eq.~(\ref{h1}).
Finally, we need an equation that determines the dynamics of $h_0(x)$.
Let us calculate the trace of Eq.~(\ref{1stab}) 
\begin{eqnarray}
{}^{(6)} R &-& \frac{3}{2}h_{|\delta}^{|\delta}
+K_a^{;a}+\frac{1}{2}h^{-1}h^{;a} K_a
=
\frac{3}{2} h 
\left( 
\omega^{;c}_{;c} + \frac{2}{3} h^{-1} h_{;c} \omega^{;c}
\right) \nonumber\\
&&+\frac{h^{-1}}{16\mbox{Im}\tau}\left|iG_{(3)}-\ast_6G_{(3)}\right|^2 
+3\kappa_{10}^2h^{-1}\left(\mu_3(y)-T_3\rho_3^{loc}\right). \label{bp}
\end{eqnarray}
Combining Eq.~(\ref{bp}) with Eq.~(\ref{tracemunu}) 
we get 
\begin{eqnarray}
{}^{(6)}R &+& \frac{1}{2} h {}^{(4)} R
-\frac{3}{2} h^{|\mu}_{|\mu} + \left( \frac{3}{2} K_{c}
-h \omega_{;c} \right)^{;c} 
= \frac{h^{-1}}{24\mbox{Im}\tau}\left|iG_{(3)}-\ast_6G_{(3)}\right|^2 \nonumber\\
&& +2 \kappa_{10}^2h^{-1}\left(\mu_3-T_3\rho_3^{loc}\right).
\label{aequation}
\end{eqnarray}
Then integrating this over the compact space we obtain the equation of motion 
for $H(x)$
\begin{eqnarray}
H_{|\delta}^{|\delta}&=&\frac{H}{3}{}^{(4)} R +
\frac{2}{3V_{(6)}}\int d^6y\sqrt{\gamma} {}^{(6)} R \nonumber\\
&&-\frac{1}{36 V_{(6)}\mbox{Im}\tau}\int d^6y\sqrt{\gamma} h^{-1}\left|iG_{(3)}-\ast_6G_{(3)}\right|^2  \nonumber\\
&&-\frac{4}{3}\frac{\kappa_{10}^2}{V_{(6)}}\int d^6y\sqrt{\gamma} h^{-1}\left(\mu_3(y)-T_3\rho_3^{loc}\right).
\end{eqnarray}
\section{\label{sec:4D}The 4D effective theory}
The effective 4D equations are summarized as 
\begin{equation}
{}^{(4)} G_{\mu\nu} =
H^{-1} \left(H_{|\mu \nu} -  g_{\mu\nu}
H^{|\delta}_{|\delta} -V g_{\mu \nu} \right),
\end{equation}
\begin{equation}
H_{|\delta}^{|\delta} =-\frac{4}{3} V + \frac{2}{3} H \frac{d V}{dH},
\end{equation}
where the potential $V(H)$ is given by
\begin{eqnarray}
\!\!\!\!\!\!V(H)&=&-\frac{1}{2V_{(6)}}\int d^6y\sqrt{\gamma} {}^{(6)} R
+\frac{1}{48 V_{(6)}\mbox{Im}\tau}\int d^6y\sqrt{\gamma}h^{-1} \left|iG_{(3)}-\ast_6G_{(3)}\right|^2 \nonumber\\
&&+\frac{\kappa_{10}^2}{V_{(6)}}\int d^6y\sqrt{\gamma} h^{-1}\left(\mu_3(y)-T_3 \rho_3^{loc}\right).
\end{eqnarray}
They can be deduced from the following 4D effective action
\begin{equation}
S_{eff}=\frac{1}{2 \kappa_4^2} \int
d^4x\sqrt{-g }\left[H{}^{(4)}R(g) -2 V(H)\right],
\label{Action}
\end{equation}
where $\kappa_4^2 = \kappa_{10}^2 /V_{(6)}$, which can be determined by integrating 
the 10D action over the six internal dimensions.

Performing the conformal transformation $g_{\mu\nu}=H^{-1}f_{\mu\nu}$
we can write the previous 4D action in the 4D Einstein frame as
\begin{equation}
S_E=\frac{1}{2 \kappa_4^2} \int
d^4x\sqrt{-f}\left[R(f_{\mu\nu})-\frac{3}{2}\left(\nabla \ln H\right)^2-2V(H)H^{-2}\right],
\end{equation}
where now $\nabla$ denotes the covariant derivative with respect to the Einstein frame
metric $f_{\mu\nu}$.
The 4D effective equations of motion in the Einstein frame are given by
\begin{equation}
R_{\mu\nu}(f)=\frac{3}{2}\left(\nabla_\mu\ln H\right)\left(\nabla_\nu\ln H\right)
+ V(H)H^{-2}f_{\mu\nu},
\end{equation}
\begin{equation}
\nabla_\alpha\nabla^\alpha\ln H=-\frac{4}{3}V(H)H^{-2}+\frac{2}{3}\frac{dV}{dH}H^{-1}.
\end{equation}
By defining $\rho(x) = i H(x)$, the kinetic term can be 
rewritten into a familiar form
\begin{equation}
S_{E, kin}= \frac{1}{2 \kappa_4^2} \int d^4 x 
\sqrt{-f} \left[R - 6 \frac{\partial_{\mu} \rho \partial^{\mu} \bar{\rho}}
{|\rho - \bar{\rho}|^2} \right].
\end{equation}
We would find the same result for the kinetic term for the universal 
K\"ahler modulus even if the wrong 
metric ansatz Eq.~(\ref{wrong}) was used to perform a dimensional reduction
where $\rho(x) = i e^{4 u(x)}$. In a region where the warping is 
negligible $h_1(y) \ll h_0(x)$, $H(x)$ can be identified as $e^{4 u(x)}$, 
which is the simple scaling of the internal metric. However, in a region 
where the warping is not negligible, the original 10D dynamics is completely 
different between (\ref{wrong}) and (\ref{metric}) \cite{HK2}.

The potential associated with the 3-form agrees with the result obtained 
in Ref.~\cite{Giddings:2005ff}. It should be emphasized that this 
potential is positive-definite. Neverthless, the 3-form contribution 
to the 4D Ricci scalar ${}^{(4)} R$ is negative definite \cite{deAlwis}
in accordance with the no-go theorem for getting de Sitter spacetime 
\cite{nogo}. The resolution
is the kinetic term of the modulus $H(x)$, that is, ${}^{(4)}R$ is not 
directly related to $V$ \cite{Giddings:2005ff}. In fact the 4D Ricci scalar
is related to the potential as 
\begin{equation}
{}^{(4)}R = H^{-1} \left(4V + 3 H^{|\delta}_{|\delta} \right) =2 \frac{dV}{dH}.   
\end{equation}
At the minimum of the potential we have ${}^{(4)} R = V= dV/dH=0$, but, 
if we move away from the minimum and the modulus $H(x)$ is moving, 
the potential cannot be read off from ${}^{(4)}R$. 

We also notice that D3 branes with $\mu_3(y)= T_3 \rho_3^{loc}$ 
do not give any gravitational energy in the 4D effective theory. 
On the other hand, anti-D3 branes with $\mu_3(y)= -T_3 \rho_3^{loc}$ 
give a potential energy. This was 
used to realize de Sitter vacuum \cite{KKLT}.

\section{Discussions}
In this letter, we presented a systematic way to derive the 4D 
effective theory by consistently solving the type IIB supergravity 
equations of motion in warped compactifications with fluxes and branes.
We used the gradient expansion method to solve the 10D equations 
of motion. The consistency condition that the integration of the 
total derivative terms over the internal 6D space vanishes gives the 
boundary conditions for the 10D gravitational fields. These 
boundary conditions give the 4D effective equations. 
Once the solutions for the 4D effective equations are obtained, 
we can determine the backreaction to the 10D geometry by solving 
Eqs.~(\ref{tracemunu}),(\ref{Sigma}) and (\ref{aequation}). 
 
In this letter, we did not introduce the stabilization mechanism 
for the modulus $H(x)$. It is essential to stabilize this universal 
K\"ahler modulus to get a viable phenomenology \cite{KKLT}. 
Usually, non-perturbative
effects are assumed to give a potential to this moduli. The non-perturbative 
effects will modify the 10D dynamics and the 4D effective
potential have to be consistent with this 10D dynamics. Most works 
so far introduce the non-perturbative potentials for the modulus filed 
directly in the 4D effective theory and it is not clear this is 
consistent with the original 10D dynamics. It is desirable to derive 
the potential for the moduli fields by consistently solving the 
10D Einstein equation with non-perturbative corrections. 
In addition, a mobile D3 brane plays a central role to realize 
inflation \cite{KKLMMT}. The D3 brane with 
$\mu_3= T_3$ probes of a no-scale compactification and the brane
can sit at any point of the compact space with no energy cost. 
In fact, the D3 brane does not give any 
gravitational energy in the 4D effective theory. However, once the 
stabilization mechanism is included, this is no longer true. 
This is in fact an important effect which generally yields a potential for 
the D3 brane that is not enough flat for slow roll. Recently, it was 
pointed out that the gravitational backreaction of the D3 brane is 
essential to calculate the corrections to the potential \cite{BDKMMM}. 

Our method can be extended to include the non-perturbative effects 
by introducing an effective 10D energy-momentum tensor 
in the 10D Einstein equations. It is 
also possible to include a moving D3 brane in our scheme 
along the line of Ref.~\cite{K2}, which studied 
the dynamics of D-branes with self-gravity in a 5D toy model.
Then we can calculate the potential for a mobile D3 brane by taking into account 
the stabilization and the backreaction of the D3 brane. In addition, it is essential 
to study how 
the motion of the D3 brane is coupled to 4D gravity in order to address 
the dynamics of inflation.  
We come back to this issue in a future publication. 

\begin{ack}

The work of KK is supported by PPARC. FA is supported by
``Funda\c{c}\~{a}o para a Ci\^{e}ncia e a Tecnologia (Portugal)", with
the fellowship's reference number: SFRH/BD/18116/2004.
\end{ack}



\begin{thebibliography}{00}
\bibitem{MAY}
L. Randall, R. Sundrum, Phys. Rev. Lett. {\bf 83}, 3370 (1999)
[hep-ph/9905221].

\bibitem{Giddings:2001yu}
  S.~B.~Giddings, S.~Kachru and J.~Polchinski,
  Phys.\ Rev.\ D {\bf 66}, 106006 (2002)
  [arXiv:hep-th/0105097].

\bibitem{KKLMMT}
  S.~Kachru, R.~Kallosh, A.~Linde, J.~M.~Maldacena, L.~McAllister and S.~P.~Trivedi,
  JCAP {\bf 0310} (2003) 013
  [arXiv:hep-th/0308055].

\bibitem{deAlwis}
  S.~P.~de Alwis,
  Phys.\ Rev.\ D {\bf 68}, 126001 (2003)
  [arXiv:hep-th/0307084];
  S.~P.~de Alwis,
  Phys.\ Lett.\ B {\bf 603} (2004) 230
  [arXiv:hep-th/0407126].

\bibitem{HK1}
H.~Kodama and K.~Uzawa, JHEP {\bf 0507}, 061 (2005)
[arXiv:hep-th/0504193].

\bibitem{Giddings:2005ff}
  S.~B.~Giddings and A.~Maharana,
  Phys.\ Rev.\ D {\bf 73}, 126003 (2006)
  [arXiv:hep-th/0507158].

\bibitem{HK2}
H.~Kodama and K.~Uzawa,
JHEP {\bf 0603}, 053 (2006)
[arXiv:hep-th/0512104].



\bibitem{gradient}
T. Wiseman, Class. Quant. Grav. {\bf 19}, 3083 (2002)
[arXiv:hep-th/0201127];\\
S. Kanno and J. Soda, Phys. Rev. D {\bf 66}, 083506 (2002)
[arXiv:hep-th/0207029];\\
T. Shiromizu and K. Koyama, Phys. Rev. D {\bf 67}, 084022 (2003)
[arXiv:hep-th/0210066];
F.~Arroja and K.~Koyama,
  Class.\ Quant.\ Grav.\  {\bf 23}, 4249 (2006)
  [arXiv:hep-th/0602068].

\bibitem{nogo}
  J.~M.~Maldacena and C.~Nunez,
  Int.\ J.\ Mod.\ Phys.\ A {\bf 16}, 822 (2001)
  [arXiv:hep-th/0007018].

\bibitem{KKLT}
  S.~Kachru, R.~Kallosh, A.~Linde and S.~P.~Trivedi,
  Phys.\ Rev.\ D {\bf 68}, 046005 (2003)
  [arXiv:hep-th/0301240].



  \bibitem{BDKMMM}
  D.~Baumann, A.~Dymarsky, I.~R.~Klebanov, J.~Maldacena, L.~McAllister and A.~Murugan,
  arXiv:hep-th/0607050.
  
  \bibitem{K2}
  K.~Koyama and K.~Koyama,
  Class.\ Quant.\ Grav.\  {\bf 22}, 3431 (2005)
  [arXiv:hep-th/0505256];
    S.~Kanno, J.~Soda and D.~Wands,
  JCAP {\bf 0508} (2005) 002
  [arXiv:hep-th/0506167].

\end{thebibliography}
\end{document}